%% file: topLHCP2014.tex
\documentclass[10pt]{article}
\usepackage{graphicx}
\usepackage{cite}

\def\Title#1{\begin{center} {\Large #1 } \end{center}}
\def\Author#1{\begin{center}{ \sc #1} \end{center}}
\def\Address#1{\begin{center}{ \it #1} \end{center}}

\newcommand\pubblock{\rightline{\begin{tabular}{l} Proceedings of the Second Annual LHCP\\ \pubnumber\\
         \pubdate  \end{tabular}}}

\newenvironment{Abstract}{\begin{quotation} \begin{center} 
             \large ABSTRACT \end{center}\bigskip 
      \begin{center}\begin{large}}{\end{large}\end{center} \end{quotation}}

\newenvironment{Presented}{\begin{quotation} \begin{center} 
             PRESENTED AT\end{center}\bigskip 
      \begin{center}\begin{large}}{\end{large}\end{center} \end{quotation}}

\def\Acknowledgements{\bigskip  \bigskip \begin{center} \begin{large}
             \bf ACKNOWLEDGEMENTS \end{large}\end{center}}

\newcommand{\beq}{\begin{equation}}
\newcommand{\eeq}{\end{equation}}
\newcommand{\bea}{\begin{eqnarray}}
\newcommand{\eea}{\end{eqnarray}}

\newcommand{\msbar}{$\overline{\mbox{MS}}$}

\input econfmacros.tex

\usepackage{color}

\newcommand\pubnumber{DESY 14-147}

\newcommand\pubdate{\today}

\def\affiliation{
  II.~Institut f\"ur Theoretische Physik\\ Universit\"at Hamburg\\
  Luruper Chaussee 149 \\
  D-22761 Hamburg, Germany}

\begin{document}

\large
\begin{titlepage}
\pubblock

\vfill
\Title{PRECISION DETERMINATION OF THE TOP-QUARK MASS}
\vfill

\Author{SVEN-OLAF MOCH}
\Address{\affiliation}
\vfill
\begin{Abstract}
Precision determinations of the top-quark mass require theory predictions with a well-defined mass parameter in a given renormalization scheme.
The top-quark's running mass in the \msbar\ scheme can be extracted with good precision 
from the total cross section at next-to-next-to-leading order in QCD.
The Monte Carlo top-quark mass parameter measured from comparison to events with top-quark decay products is not identical with the pole mass. 
Its translation to the pole mass scheme introduces an additional uncertainty of the order of 1 GeV.
\end{Abstract}
\vfill

\begin{Presented}
The Second Annual Conference\\
 on Large Hadron Collider Physics \\
Columbia University, New York, U.S.A \\ 
June 2-7, 2014
\end{Presented}
\vfill
\end{titlepage}
\def\thefootnote{\fnsymbol{footnote}}
\setcounter{footnote}{0}
%

\normalsize 


\section{Introduction}
Since the discovery of the top-quark almost 20 years ago the mass of the heaviest elementary particle currently known 
has been measured with an ever increasing and, by now, with unprecedented precision.
The top-quark mass is a fundamental parameter of the Standard Model (SM) and 
the precise value is indispensable for predictions of cross sections at the Large Hadron Collider (LHC). 
Moreover, in the absence of direct evidence for new physics beyond the SM, 
precision theory predictions confronted with precision measurements have
become an important area of research for self-consistency tests of the SM or in searching for new physics phenomena. 
This has been the motivation for significant progress, 
both on the theoretical and the experimental side, in addressing 
issues arising in precision top-quark mass determinations, see, e.g., \cite{Moch:2014tta,Juste:2013dsa} for reviews of recent activities.

\begin{figure}[ht]
  \includegraphics[width=7.25cm]{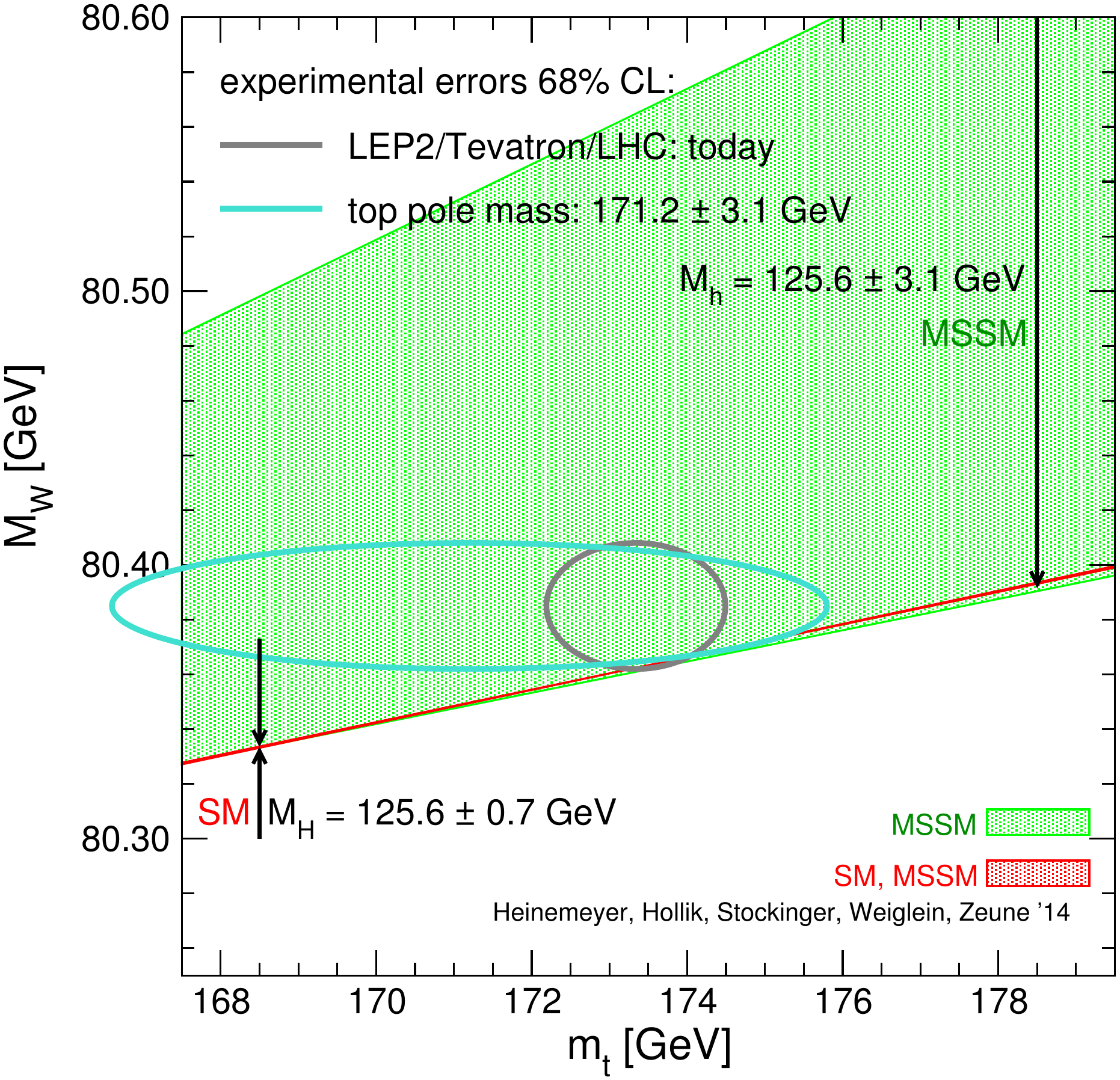}
  \hspace*{4mm}
  \includegraphics[bb = 120 450 500 760, scale = 0.65, angle=0, clip = ]{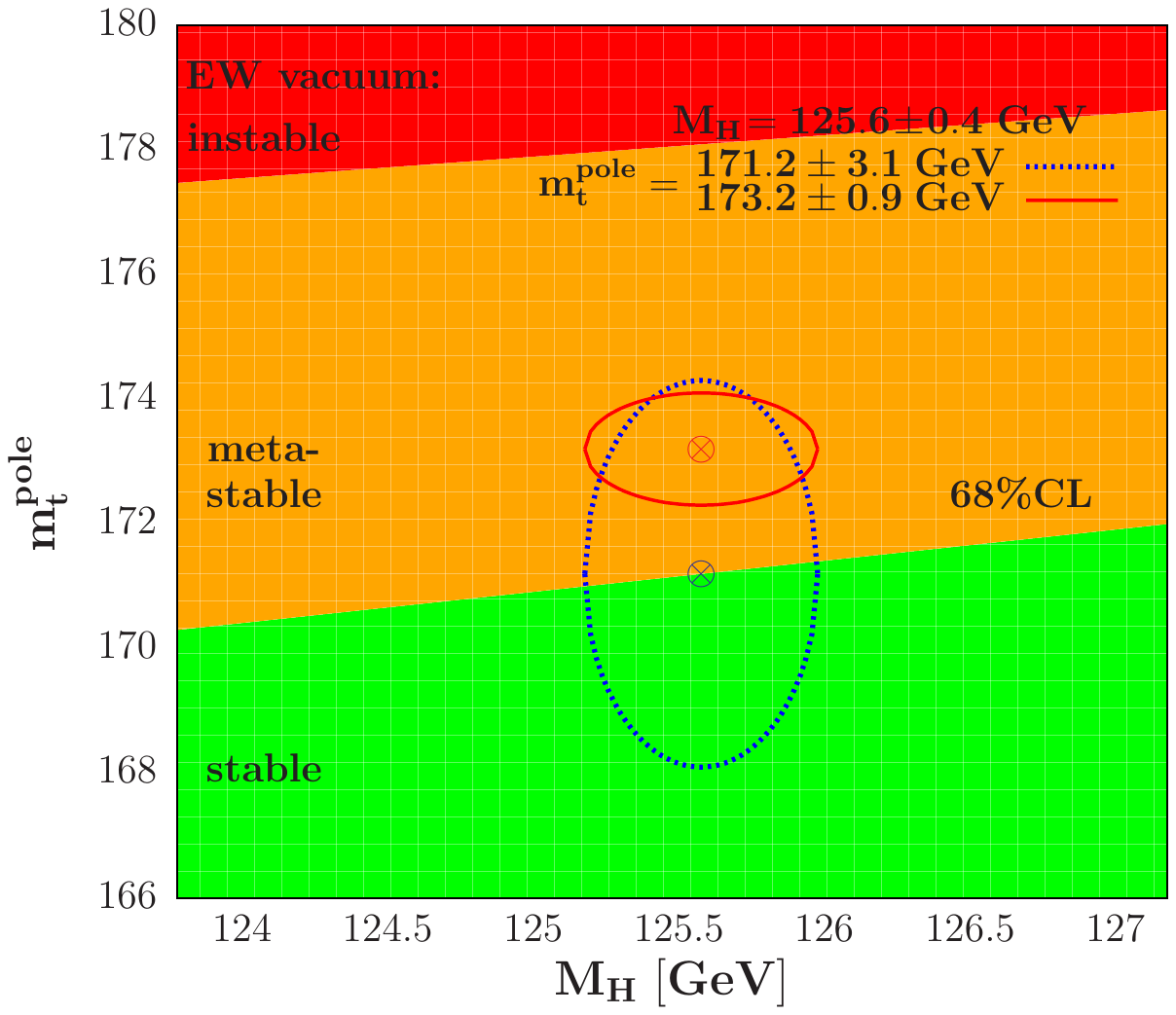}
\vspace{-4mm}
\caption{\label{fig:mtMW-mtmH}
Left: Current experimental results for $M_W$ and $m_t^{\rm pole}$ and their $1\sigma$ uncertainties in comparison with 
the SM (red band) and the MSSM prediction (light-shaded green band). 
(Figure courtesy S.Heinemeyer, cf. Ref.~\cite{Heinemeyer:2013dia}).
Right: Ellipses for the $1\sigma$ uncertainties in the [$M_H, m_t^{\rm pole}$]
plane with Higgs mass $M_H=125.6 \pm 0.4$ GeV and  $\alpha_s(M_Z)=0.1187$ 
confronted with the areas in which the SM vacuum is absolutely 
stable, meta-stable and unstable up to the Planck scale. 
(Figure from Ref.~\cite{Alekhin:2012py}).
}
\vspace*{-2mm}
\end{figure}

Here, two examples are given, where the numerical value of the top-quark mass $m_t$ directly
affects relevant physics interpretations.
On the left in Fig.~\ref{fig:mtMW-mtmH}, the current experimental results for the $W$-boson mass $M_W$ 
and the top-quark mass $m_t$ are shown in comparison with 
the theory predictions of the SM and its minimal supersymmetric extension
(MSSM) for a range of Higgs boson masses $M_H$, see, e.g.,~\cite{Heinemeyer:2013dia}.
The plot indicates consistency of the values for the various mass parameters
$M_W$, $m_t$ and $M_H$ at the level of $1\sigma$ uncertainties within the SM.
On the right in Fig.~\ref{fig:mtMW-mtmH} 
the direct impact of the top-quark mass on the Higgs sector is illustrated. 
Regions of stability of the electroweak vacuum in the $m_t$ and $M_H$ plane are plotted, 
which can be obtained from extrapolating the SM up to the Planck scale, 
see, e.g.,~\cite{Holthausen:2011aa,Bezrukov:2012sa,Degrassi:2012ry,Buttazzo:2013uya,Andreassen:2014gha}.
Thus, at high scales the existence of a well-defined minimum of the Higgs potential 
that can induce breaking of the electroweak symmetry, depends crucially on the precise numerical value of $m_t$.

\section{Top-quark mass definition}

Quark masses are not physical observables. 
This implies, first of all, that the determination of $m_t$
relies on the comparison of theory predictions $\sigma_{\rm th}(m_t)$ for cross sections 
with the experimentally measured values $\sigma_{\rm exp}$ 
for a given observable and kinematics as the best fit solution to the equation $\sigma_{\rm exp} = \sigma_{\rm th}(m_t)$. 
The accuracy of this approach is intrinsically limited by the sensitivity ${\cal S}$ 
of $\sigma_{\rm th}(m_t)$ to $m_t$, 
\begin{equation}
\label{eq:sensitivity}
  \left| \frac{\Delta \sigma}{\sigma} \right| 
  \, = \, {\cal S}  \times  \left| \frac{\Delta m_t}{m_t} \right|\, .
\end{equation}
Thus, for a given experimental error or a theoretical uncertainty $\Delta \sigma$ on the cross section, 
the greater the sensitivity ${\cal S}$ the better the accuracy for $m_t$ can be achieved.

In Quantum Chromodynamics (QCD), quark masses are simply parameters of the Lagrangian. 
They appear in the theory predictions $\sigma_{\rm th}(m_t)$ and, as such, 
they are subject to the definition of a renormalization scheme once quantum corrections at higher orders are included.
In many QCD applications the pole mass is the conventional scheme choice. 
The top-quark's pole mass $m_t^{\rm pole}$ is introduced in a gauge invariant 
and well-defined way at each finite order of perturbation theory 
as the location of the single pole in the two-point correlation function.
The pole mass scheme is, in fact, inspired by the definition of the electron mass in Quantum Electrodynamics. 
For heavy quarks, however, this has its short-comings~\cite{Bigi:1994em,Beneke:1994sw}, because 
due to confinement quarks do not appear as free particles in asymptotic states in the $S$-matrix. 
Therefore, the pole mass $m_t^{\rm pole}$ must acquire non-perturbative corrections, 
because in the full theory the quark two-point function does not display any pole.
This leads to an intrinsic uncertainty in the definition of $m_t^{\rm pole}$ 
of the order of $\Lambda_{\rm QCD}$ related to the renormalon ambiguity~\cite{Smith:1996xz}.

Fortunately, one can consider alternative definitions 
based on the (modified) minimal subtraction in the \msbar\ scheme, 
which realizes the concept of a running mass $m_t(\mu)$ at a scale $\mu$.
More generally, one can define so-called short-distance masses 
$m_t(R,\mu)$, where $R$ is a scale associated with the scheme. 
The \msbar\ mass is then just one example of a short-distance mass $m_t(R,\mu)$ with $R$ taken at the scale $R \sim m_t$. 
Other schemes define a so-called $1S$ mass~\cite{Hoang:1998hm,Hoang:1999zc} 
through the perturbative contribution to the mass of a hypothetical 
$^3S_1$ toponium bound state or a ``potential-subtracted'' (PS) mass~\cite{Beneke:1998rk}.

As alternative renormalization schemes, all short-distance masses $m_t(R,\mu)$
can be related to the pole mass $m_t^{\rm pole}$ through a perturbative series, 
\begin{equation}
\label{eq:MSRdef}
m_t^{\rm pole} = m_t^{\rm MSR}(R,\mu) + \delta m_t(R,\mu)\, , \qquad 
\delta m_t(R,\mu) = R 
\sum\limits_{n=1}^{\infty} \sum\limits_{k=0}^{n}\, a_{nk}\, \alpha_s^n(\mu)\, \ln^k \left( \frac{\mu^2}{R^2} \right)
\ ,
\end{equation}
with coefficients $a_{nk}$ known to three loops in QCD~\cite{Chetyrkin:1999qi,Melnikov:2000qh}. 

A variety of methods for top-quark mass extractions has been proposed thus far, 
see, e.g., \cite{Moch:2014tta,Juste:2013dsa}, which use a number of distinct observables. 
Examples include determinations of $m_t$ from the total cross section, 
or its extraction from the distribution of the invariant mass of a lepton and a $b$-jet, 
see, e.g.,~\cite{Beneke:2011mq,Chatrchyan:2013haa} and~\cite{Biswas:2010sa,Heinrich:2013qaa}, respectively.

With enough statistics, as expected from the LHC runs at increased collision energy, 
also exclusive observables with reconstructed top-quarks come into focus.
The (normalized) differential distribution of the $t\bar t + 1\textnormal{\normalsize -jet}$ 
cross section with respect to the invariant mass of the $t\bar t + 1\textnormal{\normalsize -jet}$ system 
displays very good sensitivity to $m_t$, ${\cal S} \sim 10 \dots 20$ in Eq.~(\ref{eq:sensitivity}) depending
on the kinematical region and can, potentially, lead to very precise values for
$m_t$, see~\cite{Alioli:2013mxa}.

All those methods employ mostly the pole mass scheme. 
The $1S$ mass and the PS mass have been considered in applications 
to hadro-production of top-quark pairs in \cite{Ahrens:2011px,Falgari:2013gwa}.
In the sequel we will discuss the determination of the running mass in the \msbar\ scheme 
and $m_t$ from reconstructed kinematics as well as the relation of those mass
parameters to the pole mass $m_t^{\rm pole}$.

%
\begin{figure}[t!]
\centerline{
  \includegraphics[width=7.45cm]{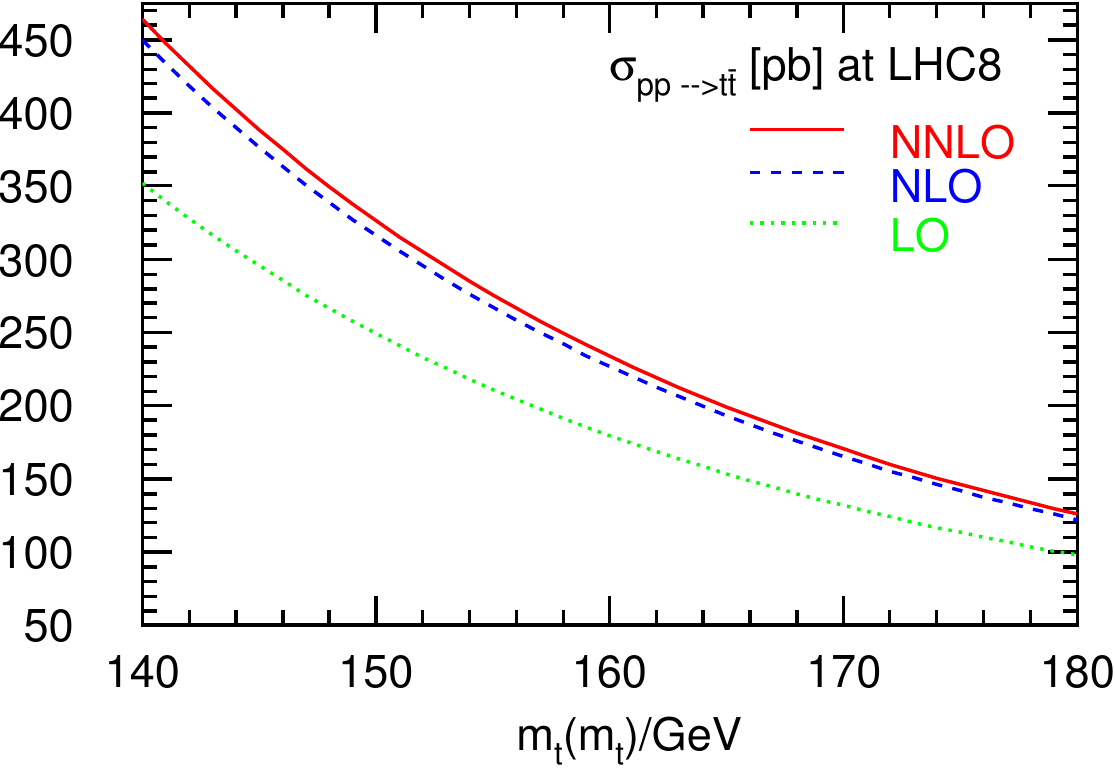}
  \hspace*{5mm}
  \includegraphics[width=7.45cm]{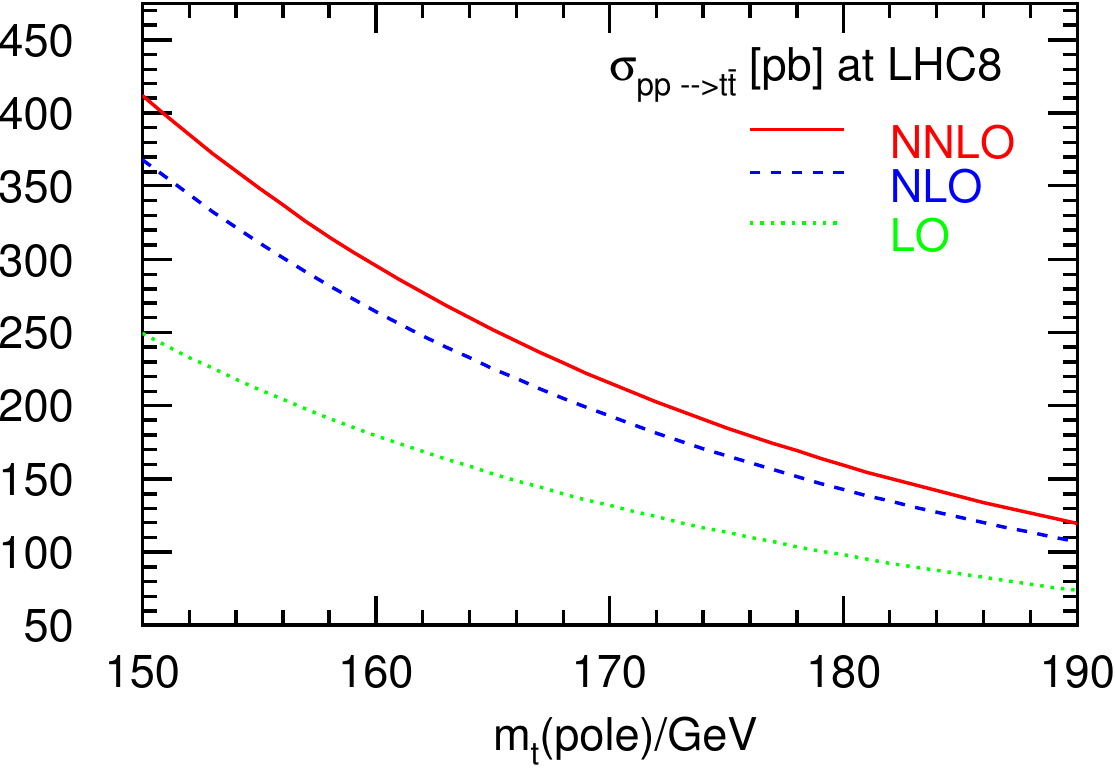}}
\vspace*{-3mm}
  \caption{\small
    \label{fig:ttbar-mass}
    The LO, NLO and NNLO QCD predictions for the 
    $t{\bar t}$ total cross section at the LHC ($\sqrt{s} = 8$~TeV) 
    as a function of the top-quark mass 
    in the \msbar\ scheme $m_t(m_t)$ at the scale $\mu = m_t(m_t)$ (left) 
    and in the on-shell scheme $m_t^{\rm pole}$ at the scale $\mu = m_t^{\rm pole}$ (right) 
    with the ABM12 PDFs. (Figure from Ref.~\cite{Alekhin:2013nda}).
  }
\vspace*{5mm}
\centerline{
  \includegraphics[width=7.45cm]{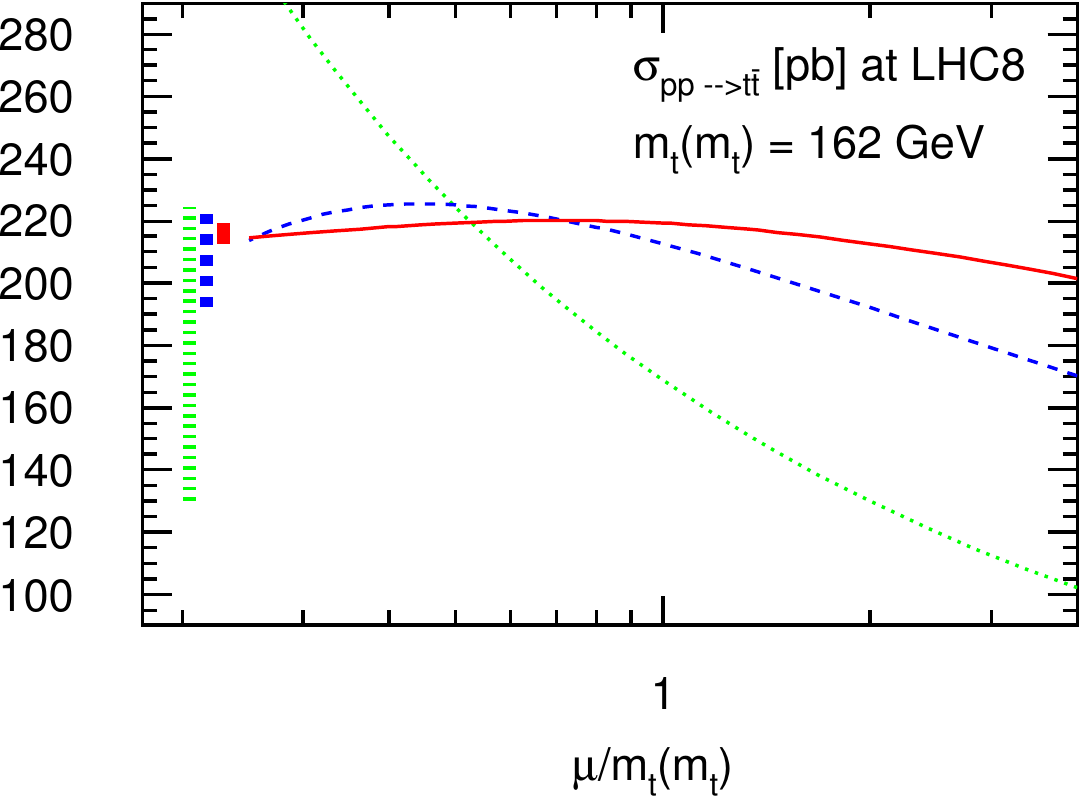}
  \hspace*{5mm}
  \includegraphics[width=7.45cm]{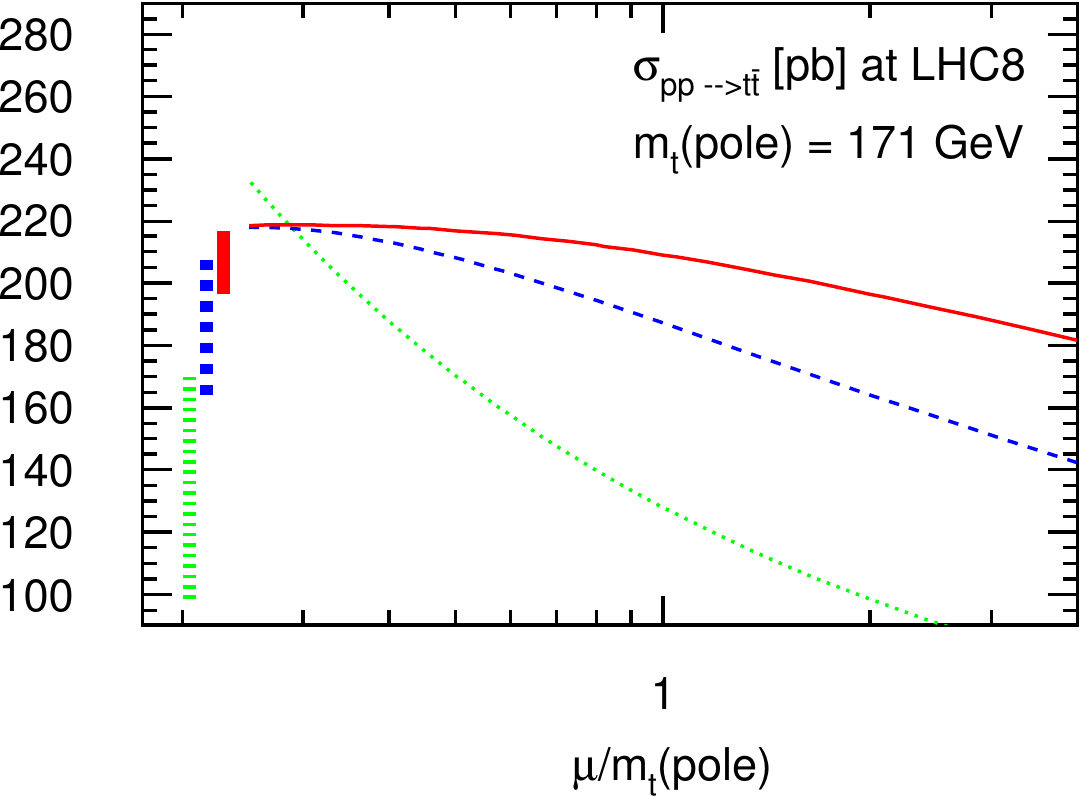}}
\vspace*{-3mm}
  \caption{\small
    \label{fig:ttbar-mu}
    The scale dependence of the LO, NLO and NNLO QCD predictions for the
    $t{\bar t}$ total cross section at the LHC ($\sqrt{s} = 8$~TeV) 
    for a top-quark mass $m_t(m_t)=162$~GeV in the \msbar\ scheme (left) 
    and $m_t^{\rm pole}=171$~GeV in the on-shell scheme (right) 
    with the ABM12 PDFs and the choice $\mu = \mu_r = \mu_f$.
    The vertical bars indicate the size of the scale variation in the standard
    range $\mu/m_t^{\rm pole} \in [1/2, 2]$ and $\mu/m_t(m_t) \in [1/2, 2]$, respectively.
    (Figure from Ref.~\cite{Alekhin:2013nda}).
}
%
\vspace*{5mm}
\centerline{
\includegraphics[width=7.45cm]{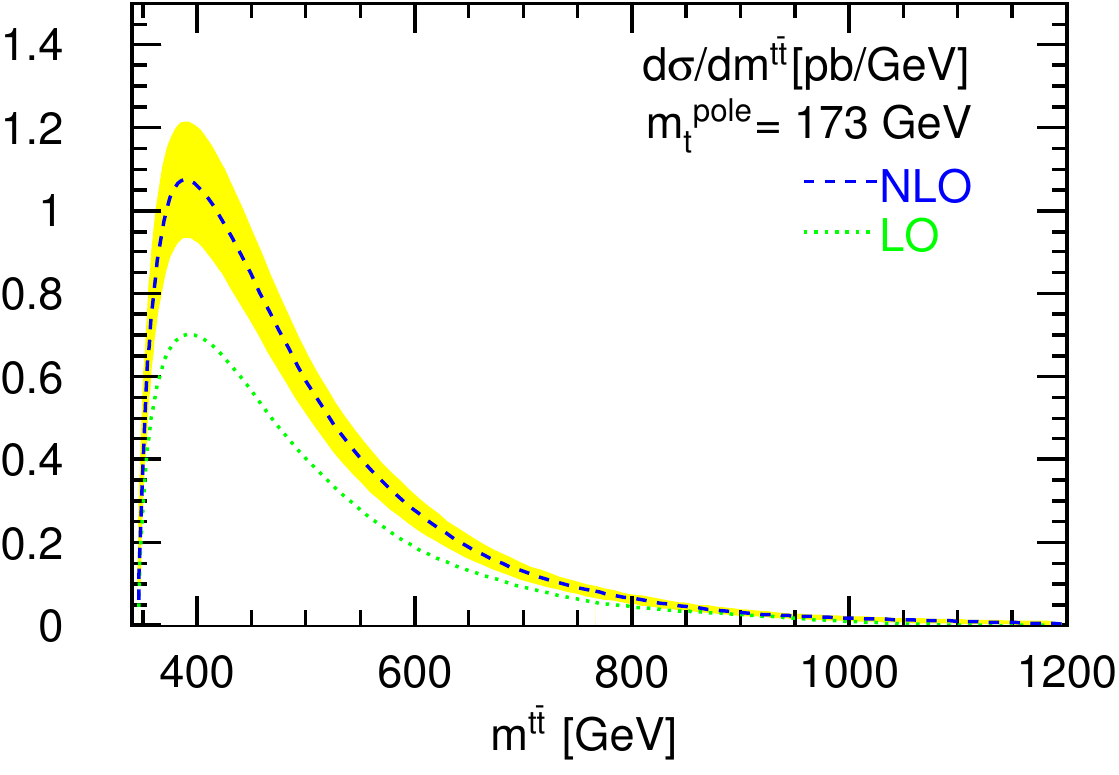}
  \hspace*{5mm}
\includegraphics[width=7.45cm]{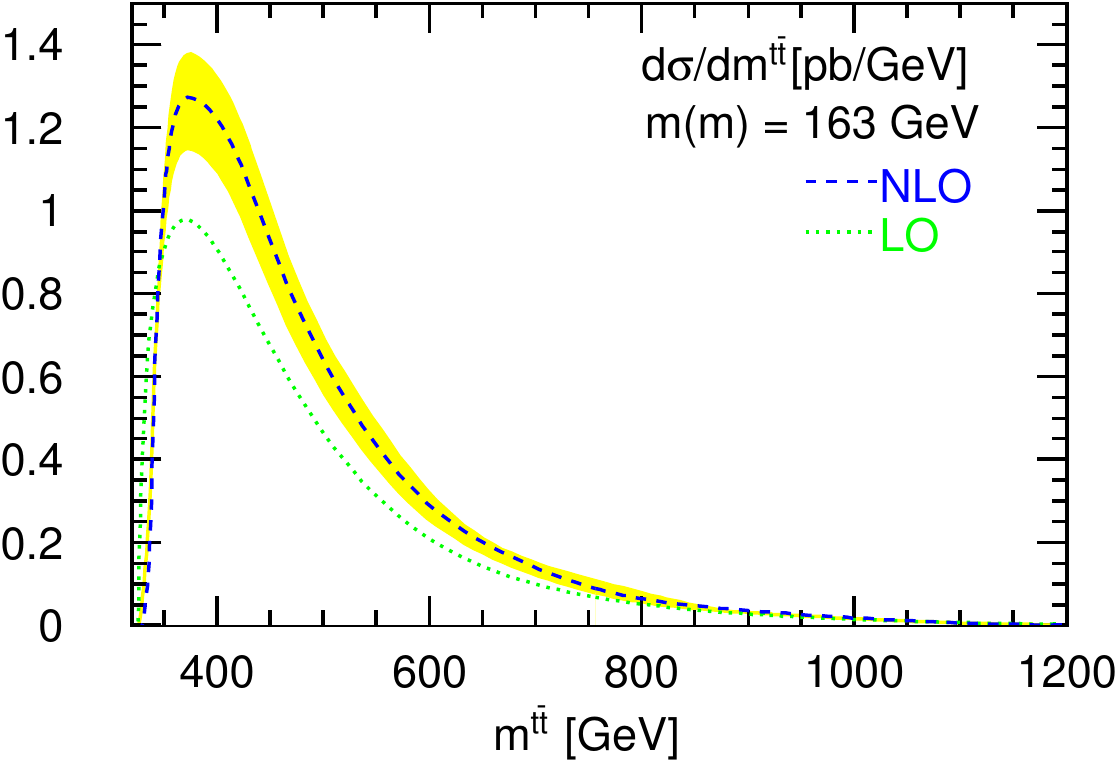}}
\vspace*{-3mm}
\caption{\small \label{fig:diff-mtt}
The differential cross section versus the invariant mass $m^{t{\bar t}}$ of the top-quark pair
in the pole (left) and the \msbar\ (right) mass scheme at the LHC
with $\sqrt{S}=8$~TeV.
The dotted (green) curves are the LO contributions, the dashed (blue) 
curves include NLO corrections and are obtained with the PDF set CT10~\cite{Gao:2013xoa}.
The scale dependence in the ranges $\mu/m_t^{\rm pole}$ 
or $\mu/m(m) \in [1/2,2]$ is shown as a band around the NLO curve.
(Figure from Ref.~\cite{Dowling:2013baa}).
} 
\end{figure}

\section{Running top-quark mass}

The running mass in the \msbar\ scheme has so far been used in theory predictions 
for the inclusive cross section~\cite{Langenfeld:2009wd,Aliev:2010zk}
or for differential distributions in \cite{Dowling:2013baa}.
Such (semi-)inclusive observables are known with good precision, i.e., 
to next-to-leading order (NLO) in perturbative QCD 
in the case of differential distributions~\cite{Campbell:2010ff,Campbell:2012uf} 
or even to next-to-next-to-leading order (NNLO) 
in the case of the inclusive cross
section~\cite{Baernreuther:2012ws,Czakon:2012zr,Czakon:2012pz,Czakon:2013goa},
see also \cite{Kidonakis:2011ca,Guzzi:2014wia,Broggio:2014yca} for approximate NNLO differential cross sections.
These computations are typically carried out in the pole mass scheme 
so that Eq.~(\ref{eq:MSRdef}) can be employed to relate $m_t^{\rm pole}$ to the \msbar\ mass.
For theory predictions in terms of the \msbar\ mass 
the perturbative expansion in the strong coupling converges significantly faster. 
At the same time, the residual scale dependence as a measure of the
remaining theoretical uncertainty is much improved when using the \msbar\ mass in contrast to the pole mass $m_t^{\rm pole}$.

These findings are illustrated in Figs.~\ref{fig:ttbar-mass} and~\ref{fig:ttbar-mu}. 
The theory predictions for inclusive top-quark pair production with the
\msbar\ and the pole mass are compared in Fig.~\ref{fig:ttbar-mass}.
The result in terms of the \msbar\ mass $m_t(m_t)$ displays 
a much improved convergence as the higher order corrections are successively added.  
The corresponding scale dependence is shown in Fig.~\ref{fig:ttbar-mu} and the predictions
with the \msbar\ mass exhibit a much better scale stability of the perturbative expansion.
It is also interesting to observe, that the  point of minimal sensitivity where 
$\sigma_{\rm LO} \simeq \sigma_{\rm NLO} \simeq \sigma_{\rm NNLO}$ is located at 
scales $\mu = {\cal O}(m_t(m_t))$, i.e., it coincides with the natural hard
scale of the process for the \msbar\ mass (Fig.~\ref{fig:ttbar-mu}, left), 
whereas it resides at fairly low scales, $\mu \simeq m_t^{\rm pole}/4 \simeq 45$~GeV
for the pole mass predictions (Fig.~\ref{fig:ttbar-mu}, right).

For the distribution in the invariant mass $m^{t{\bar t}}$ of the top
quark pair the same findings can be seen in Fig.~\ref{fig:diff-mtt}.
For the \msbar\ mass predictions the convergence is improved. 
Also the overall shape of the distribution changes in comparison to case of the pole mass, 
the peak becomes more pronounced, while the position of the peak remains stable 
against radiative corrections.
This is essential for precision determinations of the \msbar\ mass in specific kinematic regions 
of the invariant mass distribution from LHC data in the upcoming high-energy runs.

The results for the running mass imply, that experimental determinations 
of the mass parameter from the measured cross section can be performed 
with very good accuracy and a small residual theoretical uncertainty.
This has been done in~\cite{Alekhin:2013nda}, where a fully correlated fit of the running mass 
from data for the total cross section at Tevatron and the LHC 
has given the value for the \msbar\ mass at NNLO to
\begin{eqnarray}
\label{eq:moch-mt}
m_t(m_t) = 162.3 \pm 2.3~{\rm GeV}\, ,
\end{eqnarray}
with an error in $m_t(m_t)$ due the experimental data, the PDFs and the value of $\alpha_s(M_Z)$.
An additional theoretical uncertainty from the variation of the factorization
and renormalization scales in the usual range ($\mu/m_t(m_t) \in [1/2, 2]$) is small, 
$\Delta m_t(m_t) = \pm 0.7~{\rm GeV}$. 
Eq.~(\ref{eq:moch-mt}) is equivalent to the pole mass value of
\begin{eqnarray}
\label{eq:moch-mtpole}
m_t^{\rm pole} \,=\, 171.2 \pm 2.4 \pm 0.7~{\rm GeV}
\, ,
\end{eqnarray}
using the known perturbative conversion Eq.~(\ref{eq:MSRdef}) at two loops.
This is the value displayed in both plots of Fig.~\ref{fig:mtMW-mtmH}, 
which show good consistency of the procedure and also with the top-quark mass
values obtained from other determinations within the current uncertainties.
The accuracy of a mass determination in this way is limited to order 1\%, though, by the
overall sensitivity of the total cross section to the mass parameter, 
${\cal S} \sim 5$ in Eq.~(\ref{eq:sensitivity}).

%
\section{Monte Carlo mass}

The currently most precise measurement of the top-quark mass has been reported in~\cite{ATLAS:2014wva} 
as the world combination of the experiments ATLAS, CDF, CMS and D0,
\begin{equation}
  \label{eq:world-combo}
  m_t \,=\, 173.34 \,\pm\, 0.76\, \mbox{GeV} 
  \, .
\end{equation}
This combination is based on determinations of $m_t$ as a best fit to the 
mass parameter implemented in the respective Monte Carlo program used to generate the theory input. 
It is referred to as Monte Carlo (MC) top-quark mass definition 
and is, therefore, lacking a direct relation to a mass parameter in a
well-defined renormalization scheme.

Nonetheless, the MC mass definition can be translated to a theoretically well-defined short-distance mass definition at a low scale 
with an uncertainty currently estimated to be of the order of 1 GeV, see~\cite{Moch:2014tta,Hoang:2008xm}.
This translation uses the fact that multi-observable analyses like in~\cite{ATLAS:2014wva} 
effectively assign a high statistical weight to the invariant mass
distribution of the reconstructed boosted top-quarks, 
because of the large sensitivity of the system on the mass parameter, especially around the peak region.

The top-quark invariant mass distribution can be computed to higher orders in
perturbative QCD, cf., Fig.~\ref{fig:ttbar-mu}, and its peak position can also
be described in an effective theory approach based on a factorization~\cite{Fleming:2007qr,Fleming:2007xt} 
into a hard, a soft non-perturbative and a universal jet function.
Each of those functions depends in a fully coherent and transparent way on the mass at a particular scale. 
The reconstructed top object largely corresponds to the jet function which is
governed by a short-distance mass $m_t^{\rm MRS}$ at the scale of the top
quark width $\Gamma_t$, see, e.g.,~\cite{Moch:2014tta,Hoang:2008xm}.
This line of arguments allows one to systematically implement proper short-distance mass schemes 
for the description of the MC mass in Eq.~(\ref{eq:world-combo}), which can
then indeed be converted to the pole mass.

Thus, the top-quark mass parameter $m_t^{\rm MC}$ is identified with 
a scale-dependent short-distance mass $m_t^{\rm MSR}(R)$ at low scales, cf.~\cite{Hoang:2008xm},
\begin{equation}
\label{eq:MCMSR}
m_t^{\rm MC} = m_t^{\rm MRS}(3^{+6}_{-2}~\mbox{GeV})
\, ,
\end{equation}
with an uncertainty $\Delta m_t$ originating from the range of possible scales, $R \simeq 1 \dots 9~{\rm GeV}$.
The value of $\Delta m_t$ can be read off from Tab.~\ref{tab:MSRMSbar-1} as $\Delta m_t = ^{+0.32}_{-0.62}\mbox{GeV}$.
It should be emphasized, though, that this uncertainty is only an estimate 
of the conceptual uncertainty that is currently inherent in Eq.~(\ref{eq:MCMSR}). 
Very likely, the true corrections are not exactly calculable since a complete analytic control 
of the MC machinery is not feasible and the exact definition of the MC mass
also depends on details of the parton shower, the shower cut and the
hadronization model, see, e.g.,~\cite{Skands:2007zg}.

Subsequently, there are two choices to convert $m_t^{\rm MSR}$ in Eq.~(\ref{eq:MCMSR}) to the pole mass $m_t^{\rm pole}$.
The first possibility applies the renormalization group to run $m_t^{\rm MSR}$ 
from the low scales, $R \simeq 1 \dots 9~{\rm GeV}$, up to $R=m_t$ in order to obtain the
corresponding value for the \msbar\ mass $m_t(m_t)$. 
This procedure effectively resums large logarithms.
Afterwards, $m_t(m_t)$ is then converted to the pole mass at a given order in perturbation theory.
Tab.~\ref{tab:MSRMSbar-1} illustrates this procedure for 
$m_t^{\rm MSR}(3 \mbox{GeV}) = 173.40~ \mbox{GeV}$, see~\cite{Moch:2014tta} for a extensive documentation.
\begin{table}[h!]
\centering
\begin{tabular}{|c c c | c | c c c |}
\hline
$m_t^{\rm MSR}(1)$ & $m_t^{\rm MSR}(3)$ & $m_t^{\rm MSR}(9)$ & 
$m_t(m_t)$ &
$m^{\rm pole}_{\rm 1lp}$ & $m^{\rm pole}_{\rm 2lp}$ & $m^{\rm pole}_{\rm 3lp}$ \\
\hline
173.72&173.40&172.78&163.76&171.33&172.95&173.45\\
\hline
\end{tabular}
\caption{
\label{tab:MSRMSbar-1}
Columns 1-3: Top-quark MSR masses at different scales.
Column 4: \msbar\ mass $m_t(m_t)$ 
 converted at ${\cal O}(\alpha_s^3)$ for $\alpha_s(M_Z)=0.1185$ from the MSR mass $m_t^{\rm MRS}(3~\mbox{GeV})$. 
Columns 5-7: Pole masses at 1, 2 and 3 loop converted from the 
\msbar\ mass $m_t(m_t)$. All numbers are given in GeV units.
}
\end{table}

The second choice converts the short distance mass $m_t^{\rm MSR}$ at the low
scales directly to the pole mass as shown in Tab.~\ref{tab:MSRMSbar-2}.
This leads to relatively small corrections, however, the convergence of the 
perturbative expansion is poor and it is therefore disfavored.
In the application of the one-, two- or three-loop conversion formula, the value of the
mass parameter shifts by roughly $\Delta m_t \sim 0.15 \mbox{GeV}$ with every additional order.
This is due to large logarithms which need to be resummed via the renormalization group equation~\cite{Hoang:2008yj}. 
\begin{table}[h!]
\centering
\begin{tabular}{|c | c c c |}
\hline
$m_t^{\rm MSR}(3)$ & 
$m^{\rm pole}_{\rm 1lp}$ & $m^{\rm pole}_{\rm 2lp}$ & $m^{\rm pole}_{\rm 3lp}$ \\
\hline
173.40&173.72&173.87&173.98\\
\hline
\end{tabular}
\caption{
\label{tab:MSRMSbar-2}
Column 1: Top-quark MSR mass at $R=3~\mbox{GeV}$.
Columns 2-4 show the 1, 2 and 3 loop pole masses converted from the
MSR mass $m_t^{\rm MRS}(3~\mbox{GeV})$. All numbers are given in GeV units.
}
\end{table}

In summary, this leads to the following result for the pole mass, which 
corresponds to the MC mass in Eq.~(\ref{eq:world-combo}),
\begin{equation}
\label{eq:mtpole-from-MCMSR}
  m_t^{\rm pole} 
  \,=\, 173.39 \,\pm\, 0.76\, \mbox{GeV}\, \mbox{(exp)} \,+\, \Delta m_{\rm th} 
  \, ,
\end{equation}
where the small increase by $0.05\, \mbox{GeV}$ in the central value compared to Eq.~(\ref{eq:world-combo}),
is due to the shift of the three-loop pole mass with respect to $m_t^{\rm MSR}(3\, \mbox{GeV})$ in Tab.~\ref{tab:MSRMSbar-1}.
The theoretical uncertainty can be estimated to
\begin{equation}
\label{eq:Delta-mtpole}
  \Delta m_{\rm th} = ^{+0.32}_{-0.62}\, \mbox{GeV}\, (m_t^{\rm MC} \to m_t^{\rm MSR}(3\mbox{GeV})) \,+\, 0.50\, \mbox{GeV}\,  (m_t(m_t) \to m_t^{\rm pole})
\, ,
\end{equation}
where, as indicated, the first part of the uncertainty is due to the scale
choices when relating the MC mass to the short-distance mass and is subject
to the qualifications mentioned above.
The second part of the uncertainty, $\Delta m_t = +0.50\mbox{GeV}$, estimates the unknown 
higher order corrections in the conversion of the \msbar\ to the pole mass. 
Those corrections are positive and the quoted value for $\Delta m_t$ is taken as the
difference between the two-loop and the three-loop conversion, see column 6 and 7 in Tab.~\ref{tab:MSRMSbar-1}.
This part can definitely be diminished once the relation of the pole to the \msbar\ mass, 
i.e., the respective coefficients $a_{nk}$ in Eq.~(\ref{eq:MSRdef}), 
are known to four loops in QCD.

Altogether, the additional uncertainties in Eq.~(\ref{eq:Delta-mtpole}) are sizeable 
and have not been addressed in~\cite{ATLAS:2014wva} when interpreting 
the experimental measurement of the top-quark mass in Eq.~(\ref{eq:world-combo}).
The theory uncertainties are not uncorrelated, i.e., the linear sum $\Delta m_{\rm th} = ^{+0.82}_{-0.62}\mbox{GeV}$ 
in Eq.~(\ref{eq:Delta-mtpole}) should be combined in quadrature with the experimental error in Eq.~(\ref{eq:mtpole-from-MCMSR})
leading to $m_t^{\rm pole} = 173.39 ^{+1.12}_{-0.98}\mbox{GeV}$ for
the MC mass in Eq.~(\ref{eq:world-combo}).

%
\section{Summary}

\vspace*{-1mm}
\noindent

The top-quark mass is an outstanding parameter in the SM. 
Its numerical value is important for many precision tests of the model 
at current collider energies as well as for possible extrapolations to high energies. 

In QCD an unambiguous definition of the mass parameter requires the choice of
a renormalization scheme, which is conventionally taken to be the pole mass,
although this has its short-comings due to the renormalon ambiguity. 
A theoretically well-defined determination of the top-quark mass 
as a short-distance mass is possible in QCD even to NNLO 
by using inclusive observables like the total cross section for hadro-production of top-quark pairs.
This has the advantage that the theory predictions in terms of the \msbar\ mass 
converge faster at higher orders and are less affected by scale variations.
Results for the determination of the top-quark mass in this way 
have been presented in Eqs.~(\ref{eq:moch-mt}) and ~(\ref{eq:moch-mtpole}).

The top-quark mass parameter measured via kinematical reconstruction from the 
top-quark decay products by comparison to MC simulations, termed the MC mass, 
is not identical to the pole mass.
However, the measured values can be converted to the pole mass provided
certain assumption on the relation of the MC mass to a short-distance mass at
a low scale are made.
This conversion leads to an additional uncertainty of the order of 1 GeV as
quantified in Eqs.~(\ref{eq:MCMSR})-(\ref{eq:Delta-mtpole}).
Within the current accuracies, all those determinations show good consistency.
Further efforts both in theory and experiment are required though, to reduce
the uncertainty.

\Acknowledgements
This work is partially supported by Deutsche Forschungsgemeinschaft 
in Sonderforschungsbe\-reich 676 and by the European Commission through contract PITN-GA-2010-264564 ({\it LHCPhenoNet}$\,$).

\end{document}

%% file: econfmacros.tex
\def\beq{\begin{equation}}
\def\eeq#1{\label{#1}\end{equation}}
\def\eeqn{\end{equation}}

\def\beqa{\begin{eqnarray}}
\def\eeqa#1{\label{#1}\end{eqnarray}}
\def\eeqan{\end{eqnarray}}

\let\bar=\overbar

\def\Dslash{\not{\hbox{\kern-4pt $D$}}}
\def\dslash{\not{\hbox{\kern-2pt $\del$}}}

\def\msb{{\bar{\ssstyle M \kern -1pt S}}}

